\documentclass[twocolumn]{webofc}
\usepackage[varg]{txfonts}   
\begin{document}
\title{What laboratory experiments can teach us about cosmology: A chameleon example}
%
%

\author{\firstname{Clare} \lastname{Burrage}\inst{1}\fnsep\thanks{\email{clare.burrage@nottingham.ac.uk}}}

\institute{School of Physics and Astronomy,  University of Nottingham, University Park, Nottingham, NG7 2RD, UK}

\abstract{%
Laboratory experiments can shed light on theories of new physics  introduced in order to explain cosmological mysteries, including the nature of dark energy and dark matter. In this article I will focus on one particular example of this, the chameleon model.  The chameleon is an example of a theory which could modify gravity on cosmological distance scales, but its non-linear behavior means that it can also be tested with suitably designed laboratory experiments.  The aim of this overview is to present recent theoretical developments to the experimental community.  
}
\maketitle
\section{Introduction}
\label{intro}

Cosmological observations provide some of the best evidence that there must be new physics beyond the standard model of particle physics and the theory of general relativity.  Testing the theories, which could potentially provide answers to these cosmological problems is not, however, restricted to cosmological observations.  Measurements on much smaller scales, including those of laboratory experiments, can provide new information and offer the possibility to constrain or confirm new physical models. 

There is a vast number of  such theories and experiments, so in this article, after briefly reviewing the motivation for new physics coming from cosmology, I will focus on one particular example:  The chameleon model \cite{Khoury:2003rn,Khoury:2003aq}. This  is a scalar field theory, where the scalar mediates a fifth force.  As the theory is non-linear, and the effective mass of the scalar varies with the density of the environment, the fifth force is suppressed in situations traditionally used to search for new forces and modifications of gravity.  It has recently been understood, however, that laboratory experiments that probe gravitational forces with small, light particles such as atoms and neutrons can constrain the chameleon model and have the potential to exclude or detect this new particle and its associated fifth force in the near future. 

In what follows we use the mostly plus metric convention, $(-,+,+,+)$, and natural units so that $c=\hbar=k_B=1$. 

\section{Physics Beyond the Standard Model}
Cosmological observations provide convincing evidence that the standard models of particle physics and cosmology are incomplete.  This evidence comes from a wide variety of different observations that probe different distance scales, and different epochs of the evolution of the universe.  I will briefly review some of these here: 
\begin{itemize}
\item {\bf Cosmic Microwave Background.}  The cosmic microwave background (CMB) radiation is observed to be extremely isotropic across the sky.
It is very well described by a black body spectrum with a temperature of approximately $2.7 \mbox{ K}$ \cite{2009ApJ}, with temperature fluctuations of only a few hundred micro Kelvin \cite{Akrami:2018vks}.  From measurements of the amplitude and distribution of these fluctuations (as well as measurements of CMB polarization) we can learn both about the properties of the universe at very early times, and about how the universe has evolved in the time since the CMB was formed, when the universe was approximately 400,000 years old. The standard model of cosmology fits this data extremely well, however it requires: (1) That the universe contains more matter than that which we can see by the electromagnetic radiation it gives off. (2) A period of inflation, or an alternative mechanism, to explain why the CMB is so uniform (as without this assumption different parts of the sky would not have been in causal contact) and to set the initial conditions for the distribution of matter and temperature fluctuations. (3) A cosmological constant with associated energy scale approximately $2.4 \mbox{ meV}$, which would cause the expansion of the universe to accelerate at late times \cite{Ade:2015rim,Aghanim:2018eyx}.\footnote{Evidence for a cosmological constant component also comes from reconstructions of the expansion history of the universe using type 1a supernovae and baryon acoustic oscillations \cite{Betoule:2014frx}.} 
\item {\bf Galactic dynamics and gravitational lensing.} Evidence that visible matter cannot account for all of the mass in the universe also comes from galactic scales.  Including observations of the motion of stars and gas within galaxies, which is poorly fit by the assumption that their motion is determined by a gravitational force sourced by the visible mass distribution, and observations of gravitational lensing, where the photon geodesics are more lensed than can be accounted for with visible matter. For a review see Ref. \cite{RevModPhys.90.045002}. 
\item {\bf Abundance of matter and lack of anti-matter.} We observe our universe to be full of matter, and contain relatively little  anti-matter, and yet we cannot account for this imbalance within the standard models.  A process which violates B number, C and CP symmetries and occurs out of equilibrium is needed.  Within the Standard Model we do not have a sufficiently out of equilibrium process or enough CP violation. For a review see Ref. \cite{Riotto:1999yt}.
\end{itemize}
Understanding the missing matter in the universe, the very first moments in the universe's history and the asymmetry between matter and anti-matter in the universe all require physics beyond the standard models to solve them. This new physics could be new particle or forces, and the associated masses  can range from cosmologically light $\sim 10^{-42}\mbox{ GeV}$ to the GUT scale $\sim 10^{16} \mbox{ GeV}$ and beyond.  

\subsection*{The Cosmological Constant Problem}
A further puzzle is suggested by the value of the cosmological constant mentioned above. Why is the energy scale associated with the cosmological constant such a low scale?  Within a quantum theory it is not known how to protect such a small cosmological constant scale from quantum corrections, which are expected to drive the value of the constant towards the highest energy scale in the theory. Even within the standard model we would expect this to be $\sim 100 \mbox{ GeV}$ and not the $\sim \mbox{meV}$ scale that is observed. In addition we would expect the cosmological constant to  receive large contributions when the universe passes through phase transitions. Any explanation for the hierarchy between the cosmological constant and Standard Model scales must therefore explain how this value remains small throughout  cosmological history. A review of the cosmological constant problem can be found in Ref. \cite{Martin:2012bt}.  We also note here that, while current cosmological observations are well fit by a cosmological constant, possible time evolution of this `constant' is allowed by the data \cite{Betoule:2014frx,Ade:2015rim}. 

Many attempts have been made to explain the value of the cosmological constant \cite{Copeland:2006wr,Clifton:2011jh,Joyce:2014kja,Bull:2015stt}. The most common approaches are: (1) To assume that some (currently unknown) principle or symmetry sets the true value of the cosmological constant to zero.  The observed acceleration of the expansion of the universe is then due to the dynamics of a new, light field.  (2) To assume that the true value of the cosmological constant is indeed large, but that the theory of General Relativity is modified on these large scales.  The distinction between these two approaches is not as clear as it may seem at first, and there are many examples of theories which can be expressed either as a modification of gravity, or as a new matter field. For example, $f(R)$ gravity can be re-expressed as a theory of standard general relativity plus a new scalar field \cite{Brax:2008hh}. 

A common feature of these attempts to understand the value of the cosmological constant is the introduction of new, light scalar fields.  This, however, immediately introduces a new problem.  These scalars are expected to couple to standard model matter, or to the gravitational Ricci scalar (it can be shown that these two possibilities are equivalent after field redefinitions, this is known as a change of frame \cite{Will2014}).  Indeed it is difficult to forbid such couplings as they will typically be generated by quantum corrections \cite{Herranen:2015ima}. As a result the scalar field will mediate a long range fifth force with at least gravitational strength.  Laboratory experiments, and observations of the motion of bodies in the solar system have long been looking for deviations from general relativity and the presence of additional (fifth) forces without success \cite{Adelberger:2009zz}.

\section{Screening Mechanisms}
How can the tension between the theoretical  motivation for introducing new scalar fields and the precision constraints coming from local observations be alleviated? The first key point is to realize that the tight constraints of local tests apply directly only to the most simple scalar field theories with a Lagrangian of the form
\begin{equation}
\mathcal{L} = -\frac{1}{2} (\partial \phi)^2-\frac{1}{2}m^2 \phi^2 - \frac{\phi}{M}T_{\mu}^{\mu}\;,
\label{yukawa}
\end{equation}
where $m$ is the mass of the scalar, $M$ is the strength of the scalar coupling to matter, with $M\sim M_P$ for a gravitational strength coupling\footnote{The dimensionless parameter $\beta = M_P/M$ is also commonly used in the literature to denote the strength of the scalar coupling to matter. }, and $T_{\mu\nu}$ is the matter stress-energy tensor. This form of the coupling arises if matter fields move on geodesics of a metric tensor conformally rescaled by a function of the scalar field $\tilde{g}_{\mu\nu}= (1+2\phi/M)g_{\mu\nu}$. In the presence of a static, spherically symmetric source of mass $M_s$ the resulting  scalar potential felt by a unit test mass  is 
\begin{equation}
\frac{\phi(r)}{M}= -\frac{M_s}{M^2 r}e^{-mr}\;.
\label{yukawapot}
\end{equation}
The test particle feels a fifth force $\vec{F}=\vec{\nabla}\phi/M$. If the scalar is so light that it's Compton wavelength is larger than the Earth-Moon distance, then lunar laser ranging observations require $M\gtrsim 10^5 M_P$ \cite{Muller:2005sr}. 

There is no principle which forbids us including further terms in the scalar Lagrangian of Equation (\ref{yukawa}).  If we continue to require that the theory be Lorentz invariant and not possess ghost like instabilities, then additional terms will typically include higher powers of $\phi$ making the theory non-linear. As a result the scalar fluctuations mediating the fifth force will be coupled to the background, meaning that the strength of the force transmitted will depend on the environment \cite{Joyce:2014kja,Horndeski:1974wa}.\footnote{With a few exceptions, such as the introduction of a $V(\phi) = \lambda \phi^4$ potential, these additional terms in the Lagrangian will have mass dimension higher than four, and so the theory will be non-renormalisable.  Keeping quantum corrections under control in such theories  is  a challenge, but for certain theories it is understood how the theory can remain stable under such corrections \cite{Joyce:2014kja,Burrage:2016xzz}.}  

Additional terms in the Lagrangian of Equation (\ref{yukawa}), can introduce new kinetic terms, new potentials for the scalar field, or modify the coupling to matter.  The effect on the resulting scalar potential felt by a test mass is to modify Equation (\ref{yukawapot}):
\begin{equation}
\frac{\phi(r)}{M}= -\frac{f( M_s)}{M^2(\phi_{\rm bg}) r}e^{-m(\phi_{\rm bg})r}\;,
\end{equation}
where $\phi_{\rm bg}$ is the background value of the scalar field around the spherical source mass.  We see that there are three ways in which the scalar mediated potential can be suppressed compared to the form of Equation (\ref{yukawapot}): (1) The mass $m(\phi_{\rm bg})$, which is background dependent, can become large so that the scalar interactions are short ranged. (2) The coupling to matter, which may also be background dependent,  becomes weaker if $M(\phi_{\rm bg})$ becomes larger. (3) Not all of the mass of the object sources the scalar field. These effects are collectively known as screening. 

Whilst these might seem like rather baroque effects we remind the reader that a similar effect arises in electromagnetism.  A photon traveling through a plasma picks up a mass which depends on the local charge density 
\begin{equation}
m_{\rm Debye}^2=\frac{8 \pi n e^2}{ T}\;,
\end{equation}
where $n$ is the charge density of electrons, $e$ the electron charge and $T$ the temperature of the plasma.  As the charge density increases so does the effective mass of the photon and the electromagnetic interactions become increasingly short range.  One type of screening for a scalar fifth force, known as chameleon screening, works in an analogous way.  The mass of the scalar field becomes dependent on the local energy density, and so the force becomes increasingly short range in dense environments.

 In what follows we will focus on the chameleon model as an archetypal theory with screening. However we will briefly comment here about the impact of the recent observation of a neutron star - neutron star merger with both electromagnetic and gravitational waves \cite{TheLIGOScientific:2017qsa,GBM:2017lvd} on other types of screening.  This coincident observation put very tight constraints on the difference between the speed of electromagnetic and gravitational  waves. This has implications for scalar theories which screen by  introducing higher order kinetic terms (these are collectively known as Vainshtein screening).  Such kinetic terms inevitably mix the scalar with the graviton non-trivially, and a scalar background that evolves with time will result in a speed of gravitational waves which deviates from unity.  The measurement of the speed of gravitational waves  is now an important constraint for all theories which screen in this way, but particularly for theories which aim to explain the acceleration of the expansion of the universe through so called `self-acceleration' of the scalar field and zero cosmological constant \cite{Creminelli:2017sry,Crisostomi:2017lbg,Baker:2017hug,Ezquiaga:2017ekz}. We stress that the chameleon model discussed here does not modify the speed of gravitational waves, and so is not constrained by these observations. 

\section{Chameleons}
The chameleon scalar field theory requires  a simple modification of the scalar Lagrangian in equation (\ref{yukawa}).
\begin{equation}
\mathcal{L} = -\frac{1}{2} (\partial \phi)^2-V(\phi) - \frac{\phi}{M_c}T_{\mu}^{\mu}\;.
\label{cham}
\end{equation}
In the presence of non-relativistic matter $T_{\mu}^{\mu}= \rho$, where $\rho$ is the energy density, the dynamics of the scalar field are governed by an effective potential 
\begin{equation}
V_{\rm eff}(\phi)= V(\phi)+\frac{\phi \rho}{M_c}\;,
\end{equation}
and the choice of bare potential in the Lagrangian is made so that the effective potential has a minimum, and the mass of small fluctuations about that minimum increases with $\rho$.  A common choice for the potential is 
\begin{equation}
V(\phi)= \frac{\Lambda^{4+n}}{\phi^n}\;,
\label{vphi}
\end{equation}
where $n$ is an integer and $\Lambda$ a constant energy scale. The values $n=-1,-2$ are excluded as the mass of fluctuations about the minimum of the resulting effective potential does not depend on the local density, and so no chameleon behavior results. When $n<0$ only even negative integers are allowed, as otherwise the effective chameleon potential does not possess a minimum.  

For the choice of potential in Eq.~(\ref{vphi}) the minimum of the chameleon effective potential is at
\begin{equation}
\phi_{\rm min} = \left(\frac{n M_c \Lambda^{n+4}}{\rho}\right)^{1/(n+1)}\;,
\end{equation}
and the  chameleon gets an effective mass
\begin{equation}
m^2(\rho)= \frac{n(n+1)}{\Lambda^{(n+4)/(n+1)}}\left(\frac{ \rho}{n M_c}\right)^{(n+2)/(n+1)}\;.
\end{equation}

Around a static, spherically symmetric source of mass $M_s$ and radius $R_s$ the potential mediated by the chameleon scalar is
\begin{equation}
\frac{\phi(r)}{M_c}= \frac{\phi_{\rm bg}}{M_c}-\lambda_s\frac{M_s}{4 \pi M_c^2 r}e^{-m(\rho_{\rm bg})r}\;,
\end{equation}
where the coefficient $\lambda_s$ determines how much the chameleon fifth force is suppressed
\begin{equation}
\lambda_s\approx\left\{ \begin{array}{lc}
1\;, & M_s / 4 \pi R_s <M_c \phi_{\rm bg}\;,\\
4 \pi R_sM_c \phi_{\rm bg}/M_s\;, & M_s / 4 \pi R_s >M_c \phi_{\rm bg}\;.
\end{array}\right.
\label{champot}
\end{equation}
We see that for sufficiently small or diffuse objects, the chameleon force is un-screened, but for larger and denser objects $\lambda_s\ll 1$ and the force is suppressed, or screened. 

\section{Experimental searches for chameleon forces}
The screening of the chameleon potential described in equation (\ref{champot}) explains why such a force has not been detected in traditional searches for new fifth forces.  The masses in these experiments (which range from laboratory test masses to solar system bodies) are all sufficiently large that the effects of the chameleon are screened.  However, this opens up new opportunities for testing the chameleon if experimental searches can be performed with sufficiently small test masses.  It can be shown that in interesting regions of the chameleon parameter space neutrons and  atomic nuclei are sufficiently small that the chameleon force is unsuppressed. 

Precision measurements with neutrons are a perfect example of such an experiment.  As neutrons are electrically neutral they are an ideal candidate for testing gravitational physics.  Tests of gravity can be performed by allowing a beam of ultra cold neutrons to bounce above a mirror, and measuring the quantized energy levels of the neutrons.  An additional chameleon force would perturb these energy levels, and as neutrons are sufficiently small that they are un-screened, they would feel the chameleon effects without any additional suppression \cite{Ivanov:2012cb,Jenke:2014yel,Cronenberg:2015bol,Brax:2011hb}. Alternatively interferometry experiments with neutrons can also be used to test gravity.  A coherent beam of neutrons is split into two paths which would feel different chameleon effects due to the surrounding matter distribution.  When the two paths are recombined an interference fringe is produced, which is sensitive to the differential effects of the chameleon across the two paths \cite{Pokotilovski:2013pra,Brax:2013cfa,Brax:2014gja,Lemmel:2015kwa,Li:2016tux}.  

Interferometry experiments can also be performed with atoms, whilst the atomic nuclei used in these experiments are heavier and larger than neutrons, they are still small enough to be un-screened from the chameleon force.  The wave-function of the atom is split into two parts, one of which passes closer to a source mass than the other.  When the two parts of the wave-function are recombined there is a phase difference which is sensitive to the effects of the chameleon potential due to the source mass \cite{Burrage:2014oza,Burrage:2015lya,Elder:2016yxm,Hamilton:2015zga,Jaffe:2016fsh}. 

\begin{figure}[th]
\centering
\includegraphics[width=7cm,clip]{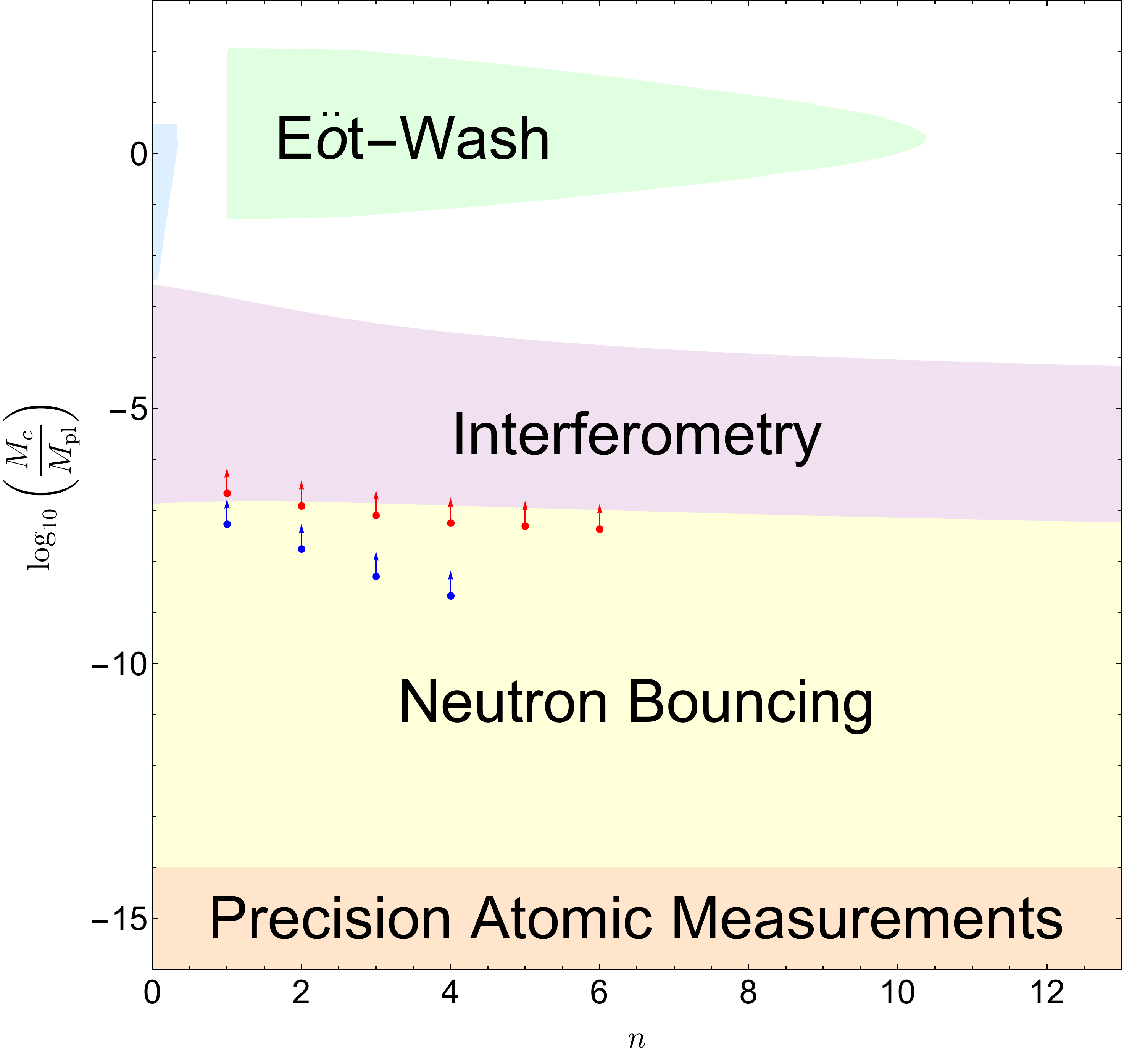}
\caption{Current bounds on the parameters $n$ and $M_c$ when $\Lambda$ is fixed to the dark energy
scale  and $n > 0$. The coloured regions are excluded.
The blue region corresponds to astrophysical tests, which includes both Cepheid and rotation
curve tests. The purple region, labeled interferometry, shows the constraints from atom interferometry. The blue and red arrows indicate the lower bounds coming from the neutron
interferometry experiments of \cite{Lemmel:2015kwa} and \cite{Li:2016tux} respectively. This figure first appeared in \cite{Burrage:2017qrf}.}
\label{fig-1}       
\end{figure}

\begin{figure}[th]
\centering
\includegraphics[width=7cm,clip]{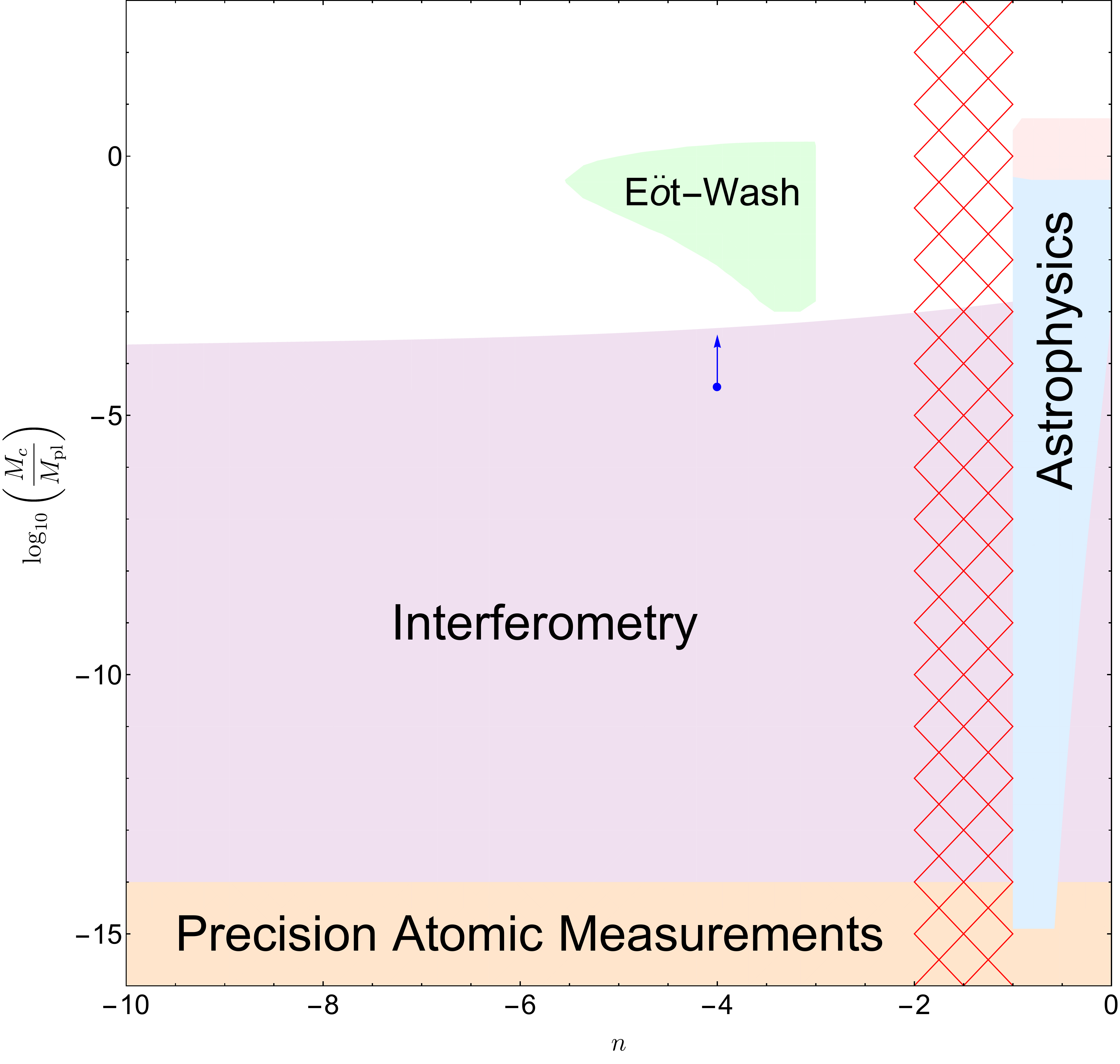}
\caption{Current bounds on the parameters $n$ and $M_c$ when $\Lambda$ is fixed to the dark energy
scale  and $n < 0$. The red hashed region indicates values of $n$ where the model is not
a chameleon, and the reader is reminded that only negative even integers are chameleons.
The coloured regions are excluded. The region labeled
astrophysics contains bounds from both Cepheid and rotation curve tests. The purple region, labeled interferometry, shows the constraints from atom interferometry. The blue
arrow indicates the lower bound coming from the neutron interferometry experiment of \cite{Lemmel:2015kwa}. This figure first appeared in \cite{Burrage:2017qrf}. }
\label{fig-2}       
\end{figure}

The constraints of both the neutron and atom experiments are shown in Figures \ref{fig-1} and \ref{fig-2}, along with other laboratory and astrophysical constraints.  The constraints are plotted for  the chameleon potential of equation (\ref{vphi}), where 
 the energy scale in the potential has been chosen to be the dark energy scale $\Lambda=2.4 \mbox{ meV}$. 

We direct readers to Reference \cite{Burrage:2017qrf} for a full review of experimental constraints on chameleon models including both laboratory based constraints and astrophysical measurements.

\section{Summary}
Understanding the observed value of the cosmological constant and the resulting accelerating expansion of the universe seems to require the introduction of new light fields into our standard models of particle physics and cosmology.  These theories are natually non-linear and include couplings between the scalar field and matter (or the Ricci scalar).  The non-linearities of the theory explain why such a scalar force has not yet been detected.  This behavior is collectively known as screening. 

The chameleon is a typical and popular screening model, in which the mass of the scalar field increases with the local density. This allows it to avoid constraints coming from searches for fifth forces performed with large and dense objects, but opens up the possibility that the fifth force could be detected in tests of gravity performed with sufficiently small objects such as atoms and neutrons. The results of these experiments impose tight new constraints on the chameleon model.  Coming improvements to the sensitivity of these measurements mean that in the near future we may be able to detect or rule out chameleons. 

\begin{thebibliography}{0}

\end{thebibliography}


\begin{thebibliography}{}
%


\bibitem{Khoury:2003aq}
  J.~Khoury and A.~Weltman,
  Phys.\ Rev.\ Lett.\  {\bf 93} (2004) 171104

\bibitem{Khoury:2003rn}
  J.~Khoury and A.~Weltman,
   Phys.\ Rev.\ D {\bf 69} (2004) 044026

\bibitem{2009ApJ}
D.~J.~Fixsen, ApJ \textbf{707}, 916-920 (2009)

\bibitem{Akrami:2018vks}
  Y.~Akrami {\it et al.} [Planck Collaboration],
  arXiv:1807.06205 [astro-ph.CO].
	
	
\bibitem{Aghanim:2018eyx}
  N.~Aghanim {\it et al.} [Planck Collaboration],
  arXiv:1807.06209 [astro-ph.CO].

\bibitem{Ade:2015rim}
  P.~A.~R.~Ade {\it et al.} [Planck Collaboration],
  Astron.\ Astrophys.\  {\bf 594} (2016) A14


\bibitem{Betoule:2014frx}
  M.~Betoule {\it et al.} [SDSS Collaboration],
  Astron.\ Astrophys.\  {\bf 568} (2014) A22
\bibitem{RevModPhys.90.045002}
Gianfranco Bertone  and Dan  Hooper,
 Rev. Mod. Phys. \textbf{90} 045002 (2018)

\bibitem{Riotto:1999yt}
  A.~Riotto and M.~Trodden,
  Ann.\ Rev.\ Nucl.\ Part.\ Sci.\  {\bf 49} (1999) 35

\bibitem{Martin:2012bt}
  J.~Martin,
  Comptes Rendus Physique {\bf 13} (2012) 566

\bibitem{Copeland:2006wr}
  E.~J.~Copeland, M.~Sami and S.~Tsujikawa,
  Int.\ J.\ Mod.\ Phys.\ D {\bf 15} (2006) 1753

\bibitem{Joyce:2014kja}
  A.~Joyce, B.~Jain, J.~Khoury and M.~Trodden,
  Phys.\ Rept.\  {\bf 568} (2015) 1
	
	\bibitem{Bull:2015stt}
  P.~Bull {\it et al.},
  Phys.\ Dark Univ.\  {\bf 12} (2016) 56
	
	\bibitem{Clifton:2011jh}
  T.~Clifton, P.~G.~Ferreira, A.~Padilla and C.~Skordis,
  Phys.\ Rept.\  {\bf 513} (2012) 1

\bibitem{Brax:2008hh}
  P.~Brax, C.~van de Bruck, A.~C.~Davis and D.~J.~Shaw,
  Phys.\ Rev.\ D {\bf 78} (2008) 104021

\bibitem{Will2014}
Clifford M.~Will, 
Living Reviews in Relativity \textbf{17} 1 (2014)

\bibitem{Herranen:2015ima}
  M.~Herranen, T.~Markkanen, S.~Nurmi and A.~Rajantie,
  Phys.\ Rev.\ Lett.\  {\bf 115} (2015) 241301

\bibitem{Adelberger:2009zz}
  E.~G.~Adelberger, J.~H.~Gundlach, B.~R.~Heckel, S.~Hoedl and S.~Schlamminger,
  Prog.\ Part.\ Nucl.\ Phys.\  {\bf 62} (2009) 102.
  doi:10.1016/j.ppnp.2008.08.002

\bibitem{Muller:2005sr}
  J.~Muller, J.~G.~Williams and S.~G.~Turyshev,
  Astrophys.\ Space Sci.\ Libr.\  {\bf 349} (2008) 457

\bibitem{Horndeski:1974wa}
  G.~W.~Horndeski,
  Int.\ J.\ Theor.\ Phys.\  {\bf 10} (1974) 363.

\bibitem{Burrage:2016xzz}
  C.~Burrage, E.~J.~Copeland and P.~Millington,
  Phys.\ Rev.\ Lett.\  {\bf 117} (2016) no.21,  211102


\bibitem{TheLIGOScientific:2017qsa}
  B.~P.~Abbott {\it et al.} [LIGO Scientific and Virgo Collaborations],
  Phys.\ Rev.\ Lett.\  {\bf 119} (2017) no.16,  161101

\bibitem{GBM:2017lvd}
  B.~P.~Abbott {\it et al.} ,
  Astrophys.\ J.\  {\bf 848} (2017) no.2,  L12

\bibitem{Crisostomi:2017lbg}
  M.~Crisostomi and K.~Koyama,
  Phys.\ Rev.\ D {\bf 97} (2018) no.2,  021301

\bibitem{Baker:2017hug}
  T.~Baker, E.~Bellini, P.~G.~Ferreira, M.~Lagos, J.~Noller and I.~Sawicki,
  Phys.\ Rev.\ Lett.\  {\bf 119} (2017) no.25,  251301
	
	\bibitem{Ezquiaga:2017ekz}
  J.~M.~Ezquiaga and M.~Zumalacárregui,
  Phys.\ Rev.\ Lett.\  {\bf 119} (2017) no.25,  251304
	
\bibitem{Creminelli:2017sry}
  P.~Creminelli and F.~Vernizzi,
  Phys.\ Rev.\ Lett.\  {\bf 119} (2017) no.25,  251302

\bibitem{Brax:2011hb}
  P.~Brax and G.~Pignol,
  Phys.\ Rev.\ Lett.\  {\bf 107} (2011) 111301

\bibitem{Cronenberg:2015bol}
  G.~Cronenberg, H.~Filter, M.~Thalhammer, T.~Jenke, H.~Abele and P.~Geltenbort,
  PoS EPS {\bf -HEP2015} (2015) 408
	
	\bibitem{Jenke:2014yel}
  T.~Jenke {\it et al.},
  Phys.\ Rev.\ Lett.\  {\bf 112} (2014) 151105
	
	\bibitem{Ivanov:2012cb}
  A.~N.~Ivanov, R.~Hollwieser, T.~Jenke, M.~Wellenzohen and H.~Abele,
  Phys.\ Rev.\ D {\bf 87} (2013) no.10,  105013

		\bibitem{Li:2016tux}
  K.~Li {\it et al.},
  Phys.\ Rev.\ D {\bf 93} (2016) no.6,  062001
	
	\bibitem{Lemmel:2015kwa}
  H.~Lemmel {\it et al.},
  Phys.\ Lett.\ B {\bf 743} (2015) 310
	
	\bibitem{Brax:2014gja}
  P.~Brax,
  Phys.\ Procedia {\bf 51} (2014) 73.
	
	\bibitem{Brax:2013cfa}
  P.~Brax, G.~Pignol and D.~Roulier,
  Phys.\ Rev.\ D {\bf 88} (2013) 083004
	
	\bibitem{Pokotilovski:2013pra}
  Y.~N.~Pokotilovski,
  Phys.\ Lett.\ B {\bf 719} (2013) 341



\bibitem{Jaffe:2016fsh}
  M.~Jaffe, P.~Haslinger, V.~Xu, P.~Hamilton, A.~Upadhye, B.~Elder, J.~Khoury and H.~Müller,
  Nature Phys.\  {\bf 13} (2017) 938
	
\bibitem{Hamilton:2015zga}
  P.~Hamilton, M.~Jaffe, P.~Haslinger, Q.~Simmons, H.~Müller and J.~Khoury,
  Science {\bf 349} (2015) 849

\bibitem{Elder:2016yxm}
  B.~Elder, J.~Khoury, P.~Haslinger, M.~Jaffe, H.~Müller and P.~Hamilton,
  Phys.\ Rev.\ D {\bf 94} (2016) no.4,  044051
			
			\bibitem{Burrage:2014oza}
  C.~Burrage, E.~J.~Copeland and E.~A.~Hinds,
  JCAP {\bf 1503} (2015) no.03,  042
	
\bibitem{Burrage:2015lya}
  C.~Burrage and E.~J.~Copeland,
  Contemp.\ Phys.\  {\bf 57} (2016) no.2,  164	
		
	
\bibitem{Burrage:2017qrf}
  C.~Burrage and J.~Sakstein,
  Living Rev.\ Rel.\  {\bf 21} (2018) no.1,  1





\end{thebibliography}
\end{document}